\documentclass[aps,prl,twocolumn,lengthcheck,superscriptaddress,showpacs,amsmath,amssymb]{revtex4}
\usepackage{graphicx}
\usepackage{dcolumn}
\usepackage{bm}

\begin{document}

\title{Dynamical instability in a trimeric chain of interacting Bose-Einstein condensates}
\author{P. Buonsante}
\affiliation{Dipartimento di Fisica and U.d.R. I.N.F.M., Politecnico di Torino, Corso Duca degli Abruzzi 24,  I-10129 Torino, ITALIA}
\author{R. Franzosi}
\affiliation{Dipartimento di Fisica, Universit\`a degli Studi di Pisa and I.N.F.N., Sezione di Pisa, Via Buonarroti 2, I-56127 Pisa, ITALIA}
\author{V. Penna}
\email[corresponding author. Mailto: ]{vittorio.penna@polito.it}
\affiliation{Dipartimento di Fisica and U.d.R. I.N.F.M., Politecnico di Torino, Corso Duca degli Abruzzi 24,  I-10129 Torino, ITALIA}
\date{\today}
\begin{abstract}
We analyze thoroughly the mean-field dynamics of a linear chain of three coupled
Bose-Einstein condensates, where both the tunneling and the central-well relative depth are adjustable parameters. Owing to its nonintegrability, entailing a complex dynamics with chaos occurrence, this system is a paradigm for longer arrays whose simplicity still allows a thorough analytical study.We identify the set of dynamics fixed points, along with the associated proper modes, and establish their stability character depending on the significant parameters. As an example of the remarkable operational value of our analysis, we point out some macroscopic effects that seem viable to experiments.
%
%
\end{abstract}


\maketitle

Since the first realizations~\cite{A:MHAnderson} 
 of Bose-Einstein condensation in atomic gases, great efforts have been aimed
at improving the control of geometrical arrangements of the condensate 
and, in particular, at realizing its fragmentation in many interacting 
components. The design of increasingly efficient trapping
schemes~\cite{A:BAnderson, A:Thomas, A:traps} has shown this program 
to be a fairly realistic perspective.
Indeed experiments involving Bose-Einstein condensates (BEC) distributed
within optical traps were successfully conducted
that provided quite large 1D  \cite{A:1D} and 2D \cite{A:Greiner} periodic arrays interacting via tunneling.
Based on magnetic trapping~\cite{A:Andrews}
the simplest case, consisting of  two coupled condensates (dimer), 
was realized as well. In parallel, the rich scenario of 
phenomena observed in BECs arrays 
(nonlinear oscillations~\cite{A:Shenoy}, 
self-trapping~\cite{A:Milburn}, 
supercurrents~\cite{A:Superf}, and solitons
~\cite{A:ASolit, A:DSolit}) raised 
a number of questions on their time evolution.
The possibility to detect and study,
both experimentally and 
theoretically, new macroscopic dynamical phenomena/effects
in BEC arrays (thus getting a deeper insight of stability 
properties and, operatively, an increased control of systems) 
has prompted an intense ongoing work.

In this perspective,
the three-site array (trimer) deserves a special attention.
According to the standard mean-field treatment, which 
is fairly satisfactory when the  average well populations are 
large~\cite{A:FPZ}, the dimer dynamics is
integrable~\cite{A:Aubry}.
The latter is described by two macroscopic complex variables
$z_i = |z_i| \exp (i \vartheta_i)$, accounting for the condensates' 
state (phase $\vartheta_i$ and population $|z_i|^2$), and exhibits two 
constants of motion, namely the total boson number and the energy.
The apparently harmless addition of a further 
coupled condensate is sufficient to make the system 
nonintegrable thus causing, in the presence of nonlinear 
BEC self-interactions, strong instabilities in extended 
regions of the phase space \cite{N:Quantum}.
That is, while keeping simple enough to be viable to a thorough
analytical study, the system displays a whole new class of behaviors which, though often overlooked, are typical of longer arrays.
Moreover, the recent achievements in the  experimental field -- and especially the control promised by micro traps \cite{A:traps} -- suggest
the realization of the trimer to be at hand. 
For all these reasons we feel that the trimer deserves a systematic analysis,  whose key results we discuss below. 

We consider an asymmetric open trimer (AOT) made of 
three coupled BECs arranged into a row, where both the 
interwell tunneling $T$ and the central-well
relative depth $w$ are adjustable parameters.
Compared with the symmetric case~\cite{CM:trimero} 
(trimer on a closed chain of equal-depth wells), the interplay
of such parameters both entails a deeper control on the dynamics
and favours the approach to experimental situations.
After recognizing the location of
fixed points in the phase space 
and the associated proper modes, we focus mostly on establishing 
via standard methods their {\it stability character} on varying 
$\tau \!= \!T/UN$ and $\nu\! =\! w/UN$, where $U$ embodies the interatomic 
scattering and $N$ is the total boson number. 
Also, we show how macroscopic 
(i.e. interesting experimentally) dynamical effects can be primed 
by selecting suitable 
 critical values of $\tau$ and $\nu$.

The essential physics of the AOT is aptly described~\cite{A:Amico, A:FPZ} 
by the Bose-Hubbard Model, which represents a gas of identical bosons
hopping across an ambient lattice.
The relevant Hamiltonian, ensuing from the boson field theory
through the {\it space-mode} approximation of field operators  
\cite{A:Jaksch}, reads 
$H\! = \! \Sigma_{k=1}^3 ( U  n_k^2-v  n_k ) - w  n_2 -
T [ a_2^+( a_1\!+\! a_3)\!+\!{\rm h.c.}]/2$,
where $v$ is the depth of wells $j= 1, 3$
($w$, $T$, $U$ are defined above),
$ n_i \!=\!  a_i^+  a_i$ counts the bosons at site $i$, and 
the destruction (creation) operators $ a_i$ ($ a_i^+$) obey 
commutators $[ a_i, a_h^+]\!=\!\delta_{i h}$.
If the well populations are not so small that a purely quantum treatment is in order, the system dynamics can be described by three variables $z_i$ within the mean-field picture. 
This is recovered via the coherent-state variational procedure of Ref.~\onlinecite{A:Amico}, where the semiclassical Hamiltonian ${\cal H} =\Sigma_{k=1}^3 (U |z_k|^4-v |z_k|^2 )  - w |z_2|^2 - T/2 [z_2^*\left(z_1+z_3\right)+{\rm c.c.}]$ is derived from the system effective action $S(t)$ ensuing in turn from the macroscopic trial state $|\Psi\rangle = \exp{(i S/\hbar)}\, \Pi_i\,|z_i\rangle$, written in terms of the Glauber's coherent states, $a_j |z_j\rangle =z_j \,|z_j\rangle $. In this framework the symbol $n_i$ is used to denote the expectation value $\langle\Psi |n_i| \Psi\rangle =|z_i|^2$.
Equipped with the Poisson brackets $\{z_j^*,z_h\}=i/\hbar\, \delta_{j h}$, 
$\cal H$  yields the dynamical equations, 
\begin{equation}
\label{E:scdyn}
\left\{
\begin{array}{l}
i\hbar \dot z_j= \left(2U|z_j|^2-v\right)z_j-\frac{T}{2}z_2 \\
i\hbar \dot z_2= \left(2U|z_2|^2-v - w\right)z_2 -
\frac{T}{2}\left(z_1+z_3\right)
\end{array}
\right.,
\end{equation}
with  $j=1,3$ in the first equation. 
Eqs. (\ref{E:scdyn}) incorporate the conservation of total boson number
$N=$ 
$ \Sigma_{j=1}^3 |z_j|^2$, since $\{N, {\cal H} \}=0$, and provide, by conjugation, the equations for $\{z^*_k\}$.
Further, 
they propagate in time the initial condition $z_1 = z_3$, 
thus revealing an integrable subregime characterized by the first 
integrals ${\cal H}$ and $N=2|z_1|^2+|z_2|^2$: We refer to such a
regime as the {\it dimeric} regime (DR), as opposed to the 
{\it nondimeric} regime (NR) where $z_1 \neq z_3$.

The distinctive features characterizing a given Hamiltonian 
dynamics are deduced by exploring its phase-space structure. 
This is attained in the first place by working out the location and 
the local character of its fixed points.
For Eqs. (\ref{E:scdyn}), the latter are issued from
\begin{equation}
\label{E:fpsys}
\left\{
\begin{array}{ll}
0\!=\! \left(2U|z_j|^2-\mu\right)z_j-\frac{T}{2}z_2 & j\!=\!1,3\\
0\!=\! \left(2U|z_2|^2-\mu - w\right)z_2 \!-\! \frac{T}{2}\left(z_1\!+\!z_3\right) &
\end{array}
\right.
\end{equation}
where $\mu=v+\chi$ and $\chi$ is a Lagrange multiplier 
selecting the conserved value of $N$.
Every solution $(\eta_1,\eta_2,\eta_3 )$ of Eqs. (\ref{E:fpsys}) 
naturally provides a periodic solution of Eqs. (\ref{E:scdyn}) 
of the form $z_j(t) = \eta_j\,\exp[i\,\chi\,t/\hbar]$. 
The global phase symmetry $z_j \mapsto z_j \,\exp[i\, \Phi]$ 
($ \Phi \in [0,2\,\pi]$) allows one to replace $z_i$ with 
$x_i \in \mathbb{R}$ thus reducing system (\ref{E:fpsys}) 
to
\begin{equation}
\label{E:Rfpsys}
\left\{
\begin{array}{ll}
0\!=\! (2 \,x_j^2-m) x_j-\frac{\tau}{2}\,x_2 & j=1,3\\
0\!=\! \left(2\,x_2^2-m \!-\! \nu\right)x_2 -\frac{\tau}{2}\,\left(x_1+x_3\right)
\end{array}
\right.
\end{equation}
where  $m = \mu/UN$, and ${\bf x}^t\!=\!(x_1,x_2,x_3)$ is such that
$z_j \!=\! \sqrt{N} x_j \exp[i\,\Phi]$ and $\Sigma_i x_i^2 \!=\!\Sigma_i n_i/N\!=\! 1$.
Notice that the structure of system (\ref{E:Rfpsys}) 
entails that the solution where either one or both of the 
peripheral sites are depleted is trivial:
 $x_{1,3}=0 \Rightarrow {\mathbf x} \!=\!{\mathbf 0}$.
The situation where the central well is depleted, henceforth referred 
to as {\it central-depleted well} (CDW)
yields $x_1=$ $-x_3 = \pm \sqrt{m/2}=\pm 1/\sqrt 2$. Note that this represents
the trimeric counterpart of dimer $\pi$-states \cite{A:Shenoy}.
In general, except for the simple CDW case, solving 
system (\ref{E:Rfpsys}) for $\tau\!\neq\!0$ is found to be equivalent to finding 
the real roots of the quartic polynomial in ${\alpha}$ \cite{N:polycoefs}
\begin{equation}
\label{E:poly}
{\cal P}_{\tau \nu}({\alpha}) = 
{\alpha}^4 + b_{\tau \nu} {\alpha}^3 + c_{\tau \nu} {\alpha} + d =0,  
\end{equation}
Once a solution is found in terms of ${\alpha}$, the 
relevant configuration ${\bf x}$ is recovered as $x_2\! = \! {\alpha}R({\alpha})/\sqrt{1\!+\!{\alpha}^2}$,
\begin{equation}
x_{1,3}=(X_1\!\pm\! X_3)/\sqrt 2\;\;({\rm NR}),\quad x_{1,3}=X_{1,3}\;\;({\rm DR})
\label{COMP}
\end{equation}
where 
$X_3\! =\!(1-X_1^2-x_2^2)^{\frac{1}{2}}$,
$X_1\! =\! R({\alpha})/\sqrt{1+{\alpha}^2}$ and $R({\alpha}) =\!(1-\tau\,{\alpha}/\sqrt 8)^{\frac{1}{2}}$ in the NR, while $X_1 = R({\alpha})/\sqrt{2\,(1+{\alpha}^2)}$, $R({\alpha}) = 1$ in the DR.

\begin{figure}
\begin{center}
\includegraphics[width=8.5cm]{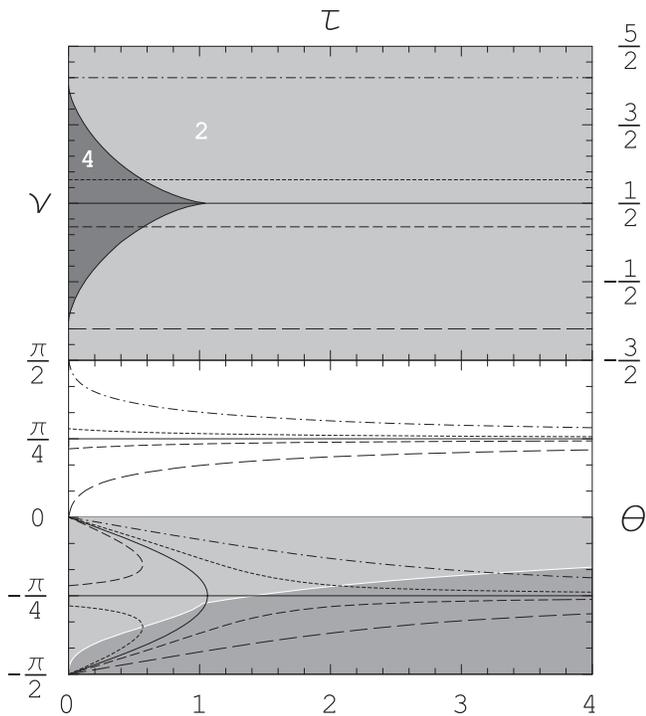}
\caption{\label{F:dim} Stability diagram for the DR. Upper panel: number of fixed points. Lower panel: character of the fixed points. The color keys of both panels, sharing the same $\tau$ axis,  are the same as in the GR case (see caption of Fig.~\ref{F:gen}).}
\end{center}
\end{figure}

\begin{figure}
\begin{center}
\includegraphics[width=8.5cm]{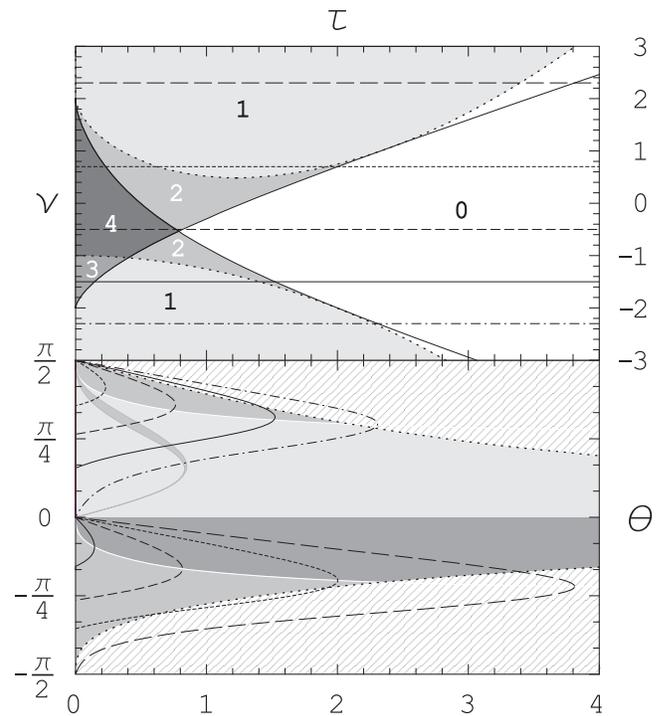}
\caption{\label{F:gen}  Stability diagram for the NR. Upper panel: number of fixed points. The darker is a region, the larger is the relevant number of solutions, also displayed. The lines at a fixed value of $\nu$ correspond to the curves $\tau_\nu(\theta)$ appearing in the lower panel with the same dashing style. Lower panel: character of the fixed points. Dark gray: maxima; Medium gray: unstable saddles; Light Gray: stable saddles (only in the NR); White: minima (only in the DR); Diagonal lines: forbidden regions (only in the NR). Both panels share the same $\tau$ axis.}
\end{center}
\end{figure}

The study of Eq. (\ref{E:poly}) allows one to construct the 
upper-panel diagrams of Figs.~\ref{F:dim} and \ref{F:gen},
where the same shade of gray characterizes regions of the 
$\tau \nu$ plane featuring the same number of fixed points.
Also, it represents an essential element in building the 
{\it stability diagrams}
[displayed in the lower panel 
of Fig.~\ref{F:dim} (Fig. \ref{F:gen})
for the DR (NR)] 
which are the central result of this
letter.
Operationally, such diagrams, depicted in the $\theta \tau$ plane 
(where $\theta=\arctan({\alpha})\! \in [-\pi/2,\pi/2]$), allow 
to determine both the location and the stability character of
the fixed points.   
For a given parameter pair $(\tilde \tau,\tilde \nu)$, the latter 
are obtained by intersecting the straight line 
$\tau = \tilde \tau$ and the curve $\tau= \tau_{\tilde \nu} (\theta)$,
where $\tau =$ $\tau_{\nu} (\theta)$ is implicitly defined by equation 
${\cal P}_{\tau \nu}({\alpha})=0$. The fixed-point components $x_i$'s 
are obtained from the relevant $\theta$ via Eqs. (\ref{COMP}).
The lower panel of Fig.~\ref{F:dim} (\ref{F:gen})
displays the graph of $\tau=$ $\tau_{\nu} (\theta)$ 
for five (four) choices of $\nu$, which we
discuss below. We remark that such diagrams embody the whole 
information on trimer dynamics. There, the fixed points are
identified as maxima, minima and (stable/unstable) saddles 
based on the shade of gray of the region they lie in.

{\bf Fixed-points local character}. This is identified analytically 
through a standard study of the signature of the quadratic form 
associated to a given Hamiltonian, along the lines 
of Ref.~\onlinecite{CM:trimero}.
Locally, to the second order in the displacements 
$\xi_j\! =\!\sqrt{N}\left( q_j + i \, p_j\right) = 
z_j \!-\! \zeta_j$ around the fixed point
$\zeta_j = \sqrt{N} x_j$,
one gets ${\cal H}(z_j) - \chi\, N \!= \! {\cal H}(\zeta_j) +
U \, N^2 \tau \, h({\bf q},{\bf p})/2$, 
with ${\bf q}^t \!=\! (q_1, q_2, q_3)$, ${\bf p}^t\! =\! (p_1, p_2, p_3)$ and 
$h({\bf q},{\bf p}) =  {\bf q}^t Q {\bf q} +{\bf p}^t P {\bf p}$.
Remarkably, the six-by-six matrix relevant to the quadratic form 
$h({\bf q},{\bf p})$ in the vicinity of a fixed point, 
{\it separates} into two three-by-three symmetric submatrices $P$, $Q$
depending on the $x_j$'s, $\tau$ and $\nu$.
Also,
when the displacements' constraint
${\bf x}^t {\bf q} = 0$ (issued from $\Sigma_j |z_j|^2 \equiv N$) 
is accounted for, the rank of $Q$ further reduces to two. Likewise 
only two eigenvalues of $P$ affect the fixed-point character,
the third one being identically zero. 
Indeed it is easy to check that the relevant  eigenvector 
is ${\bf p}_0 \!\propto\! {\bf x}$.
The non-zero eigenvalues of $P$ have opposite signs in the CDW 
configuration (which is therefore a saddle), and the signs of 
$x_1/x_2$ and $x_3/x_2$ otherwise. 
Hence the character of a fixed point depends on the signs 
of the two pairs of significant eigenvalues of $P$ and $Q$.

{\bf Linear stability analysis}.
The linear stability \cite{B:Ott} of the fixed point can be determined 
by studying the evolution of the displacements ${\bf q}$ and 
${\bf p}$. This is governed by the differential equations 
$\dot {\bf q}= \{{\bf q},{\cal H}\} =  \sigma \,P \, {\bf p}$ 
and $\dot {\bf p} = \{{\bf p},{\cal H}\} = 
-\sigma \,Q \, {\bf q}$  (with $\sigma = {U N \tau}/{2\hbar}$), 
through the Poisson brackets $\{q_j,p_k\}=\delta_{j k}/(2\,N\hbar)$ 
stemming 
from the original ones $\{z^*_j,z_k\}=i\delta_{j,k}/\hbar$.
Such ${\bf qp}$ equations provide 
the dynamical six-by-six 
matrix $\bf M_{6}$ with diagonal (three-by-three) blocks vanishing, 
and off-diagonal blocks that coincide with matrices $P$ and $-Q$~\cite{N:M6b}.

Based on the standard criteria,
a fixed point is (linearly) stable when none of the six complex 
eigenvalues $\{\lambda_j\}_{j=1}^6$ inherent in $\bf M_{6}$ features 
a positive real part. Otherwise the fixed point is 
{\it unstable} and a chaotic behavior may arise. 
The $\lambda_i$'s are conveniently obtained as the square roots of the   
eigenvalues $\{\Lambda_j\}_{j=1}^3$ of $-PQ$. Indeed, due to the block off-diagonal form of $\bf M_{6}$ and to the symmetry of $P$ and $Q$, $\det(\lambda\! -\!{\bf M_{6}})\!=\!\det(\lambda^2 \!+\!PQ)$. 
As expected, this implies that maxima and minima are stable points. 
Indeed for such configurations the eigenvalues of $P$ and $Q$ have the same 
signs (non-positive for the maxima and non-negative for the minima), yielding 
therefore non-positive $\Lambda_j$'s and purely imaginary $\lambda_j$'s. 
Further, the fact that one of the eigenvalues of $P$ is zero implies that one of the  $\Lambda_j$'s, and hence two of the $\lambda_j$'s, are zero as well. 
The four remaining $\lambda_j$'s are the roots 
of a biquadratic polynomial of the form $\lambda^4 \!-\!s \lambda^2 \!+\! p$. 
A fixed point is therefore stable if the conditions 
$s^2-4 p<0$, $p\! >\!0$ and $s\!<\!0$ are simultaneously met.
%

{\bf Discussion}.
The character/stability of fixed points are fully described by the 
lower panel of Fig.~\ref{F:dim} (Fig.~\ref{F:gen}) --referred to 
the DR (NR)-- through the graphs of function $\tau_{\nu} (\theta)$ 
for five (four) significant choices of $\nu$, distinguished by 
different dashing patterns.
In the figure upper panels a straight horizontal
line featuring the same dashing style as the corresponding 
$\tau_{\nu} (\theta)$ curve allows one to read the relevant
value of $\nu$. As noted above, each pair $(\bar \nu,\bar \tau)$
selects the set
$\{ \theta_r :
{\bar \tau} =\tau_{\bar \nu}(\theta_r ),\, r \le 4 \}$
giving the fixed-point components.
Depending on the shade of gray filling the region it lies in, 
a fixed point is either a minimum (white), a stable/unstable saddle 
point (light/medium gray) or a maximum (dark gray).
In the DR
(Fig.~\ref{F:dim}) each curve $\tau_\nu (\theta)$ 
has two branches featuring asymptotes at $\theta=\pm \pi/4$. 
For $-1\leq \nu \leq 2$, a third bell-shaped branch 
[see $\tau_{\nu} (\theta)$ for $\nu = 0.2, 0.4$] crops up. 
Hence the DR always has two fixed points, of which 
one is always a minimum, whereas the other is a maximum 
(unstable saddle) for large (small) $\tau$'s.
For $\nu=1/2$ the unbounded branches actually collapse on their 
asymptotes, thus providing $\tau$-independent solutions \cite{N:nu05}.
In the NR, Fig.~\ref{F:gen}, $\tau_\nu (\theta)$ features one or two 
bell-shaped branches, depending on whether $|\nu|\ge 2$
or $|\nu|<2$, respectively. Notice that in some cases entire portions
of such branches, namely the ones lying within the patterned regions, must be 
discarded since, despite they are real, the relevant roots of polynomial 
(\ref{E:poly}) yield complex solutions of system (\ref{E:Rfpsys}).
\begin{figure}
\begin{center}
\vskip 0.2cm
\includegraphics[width=8.5cm]{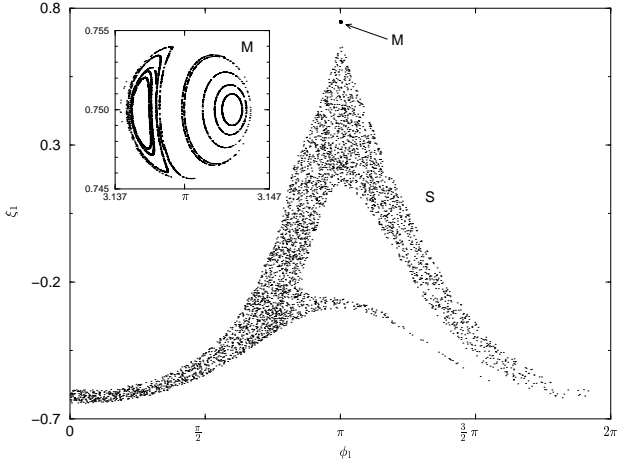}
\caption{\label{F:Psec} 
Sets S and M of Poincar\'e sections of trajectories 
based close to $C_0$ for two choices 
of significant parameters $\tau,\nu$ (see the text for further detail).
Sections are cut at $\xi_2=-3/4$ within the reduced phase space 
$(\xi_1, \xi_2, \phi_1,\phi_2)$, 
[with $\xi_1 \!= \!(n_3-n_1+n_2)/N$, $\xi_2= (n_3-n_1-n_2)/N$] 
ensuing from $\dot N= 0$ (see also Ref. \onlinecite{CM:trimero}).}
\end{center}
\end{figure}
The dotted curves appearing in the upper panel of Fig.~\ref{F:gen} 
are actually the counterparts of the dotted curves delimiting such 
forbidden regions, and significantly modify the simpler picture 
ensuing from the mere study of the real roots of polynomial
(\ref{E:poly}) (solid black lines).
For $\tau < 2 [2/3^{1/2} - 1]^{1/2}$ and $\nu$ ranging around 
$\nu =2(3^{1/2} -2)$ a  four-solution lobe is found.
As observed above, 
maxima and minima are predicted to be stable fixed points.
Noticeably, the stability analysis evidences that in the NR a 
significant fraction of the saddle points is dynamically stable
(light gray), while in the DR the saddle points are always
unstable (medium gray).
Also, while Fig.~\ref{F:dim} shows how 
the minimum always belongs to the DR, the absolute maximum is 
found either in the DR or in the NR depending on the values of $\nu$, 
$\tau$. As to the CDW saddle, it is unstable in the region
$\{\nu \!< \!-1,\tau \!\leq\! \sqrt{-1-\nu} \}\cup \{|\nu|<1,\tau \geq [(1+\nu)^3/(2-2\,\nu)]^\frac{1}{2}\}$ 
 and stable elsewhere.

As already mentioned, the versatility of
micro traps \cite{A:traps} promises a full control on
the features of the trapping potentials, and hence on the
parameters $\tau$ and $\nu$. It is worth noting that
present experimental setups \cite{A:1D,A:Greiner} already allow
to tune the ratio $T/U$ so that, according to
the estimates in Ref.~\onlinecite{A:Jaksch}, $\tau$ spans the range 
displayed in Figs.~\ref{F:dim} and \ref{F:gen} if a macroscopic 
$N=10^3\div 10^5$ is considered \cite{A:lungo}.

We conclude by emphasizing
two interesting features of AOT, 
among the many suggested by our analysis, which seem to be
more readily apt to experimental tests.
As to the just mentioned CDW saddle we recall that its form
($n_2=0$, counter-phased side wells with $n_1=n_3$) does not 
depend on parameters $\tau$ and $\nu$.
Hence a system prepared in such a configuration with
a weak ($\tau \ll 1$) interwell tunneling, would remain
in such apparently unnatural state also when tunneling is 
enhanced ($\tau\! \gg\! 0$), provided the central well is sufficiently 
deep ($\nu > 1$) to ensure the dynamical stability of the fixed point. 
Notice that preparing two separate counterphased condensates is by no means an easy task. In this respect the above described phenomenology suggests an operational method to check if the system is satisfactorily close to the desired configuration. 
%
Further, we remark that, owing to the relatively simple algebraic structure 
of the dimeric equations, there exists a simple, actually linear, functional 
relation among $\tau$ and $\nu$ ensuring that a fixed point is characterized 
by the same coordinates $x_i$'s independent of the parameters \cite{N:radius}. 
This allows one to evaluate with little effort different sets of parameters 
$(\tau,\nu)$ for which the same fixed point belongs to regions having
a different stability character. Then, a relatively simple tuning of 
$\tau$ and $\nu$ leads to change the fixed-point character from stable 
to unstable, thus inducing chaos onset.
Such an effect is manifest in Fig.~\ref{F:Psec},
where two sets (labeled by $M$ and $S$) of Poincar\'e sections, 
both issued from initial conditions quite close to the 
configuration $C_0$ ($n_1 = n_3 = N/8$, $n_2 = 3/4 N$,
$\phi_1 =$ $ \vartheta_2 -\vartheta_1= \pi$, $\phi_2 =$ $\vartheta_3 -\vartheta_2=$ 
$ -\pi$) are plotted for $(\tau_{\rm M},\nu_{\rm M})=(7/[4 \sqrt 6],2/3)$ 
and $(\tau_{\rm S},\nu_{\rm S})=(13/[8 \sqrt 6],17/24)$,
so that the corresponding points $(\tau_{{\rm M},{\rm S}},\theta=-\pi/3)$
lie just in the maxima and the saddle regions, 
respectively.
Notice indeed that the former parameter choice yields the 
regular section expected for a stable point. Conversely, 
when the fixed point is a saddle, 
trajectories based very close to it invade a large phase-space region, 
as testified by Fig.~\ref{F:Psec} where the 
relevant Poincar\'e sections are undistinguishably merged into a fuzzy cloud of points.
The analytical study exposed in the present letter is widely confirmed 
by numerical simulations where the chaoticity of the unstable fixed 
points is manifest.
A thorough discussion of the
AOT phenomenology and of further macroscopic effects disclosed by 
the stability diagrams will be presented in a later paper~\cite{A:lungo}.

\bibliographystyle{etal2}


\begin{thebibliography}{10}
\bibitem{N:nu05}
This is related to the peculiar symmetry of system (\ref{E:Rfpsys}),
  characterizing the DR ($x_1=x_3$): $\{(\bar x_1,x_2); \tilde \nu\}
  \leftrightarrow \{(\bar x_1,x_2); -\tilde \nu\}$, where $\tilde \nu =
  \nu-1/2$ and $\bar x_1 = \sqrt 2 x$.
\bibitem{N:radius}
In the generic regime such functional relation is complicated essentially by
  the dependance of $R$ on $\theta$ and $\tau$.
\end{thebibliography}
\end{document}